\documentclass[11pt]{article}

\usepackage{amsmath, euscript, amssymb, amsfonts}
\usepackage[matrix,arrow,curve]{xy}

 \newcommand{\supp}{\mathop{\rm supp}\nolimits}

 \newcommand{\hotimes}{\mathbin{\hat{\otimes}}}



\newcommand{\oR}{{\mathbb R}}
\newcommand{\oV}{{\mathbb V}}
\newcommand{\oC}{{\mathbb C}}
\newcommand{\oW}{{\mathbb W}}

\newcommand{\defeq}{\overset{\text{def}}{=}}

\begin{document}

\thispagestyle{empty}

\baselineskip=15pt

\bigskip

\phantom{X}

\vspace{2cm}

\begin{center}
{\Large\bf Reconstruction in quantum field theory with a
fundamental length }

\medskip
Dedicated to the memory of Professor V.~Ya.~Fainberg

\medskip

\bigskip

{\large M.~A.~Soloviev}\footnote{E-mail: soloviev@lpi.ru}
\end{center}

\vspace{-0.1cm}

 \centerline{\sl P.~N.~Lebedev Physical Institute}
 \centerline{\sl Russian Academy of Sciences}
 \centerline{\sl  Leninsky Prospect 53, Moscow 119991, Russia}

\vskip 3em

\begin{abstract}

In this paper, we establish an analog of Wightman's reconstruction
theorem for nonlocal quantum field theory with a fundamental
length. In our setting, the Wightman generalized functions are
defined on test functions analytic in a complex
$\ell$-neighborhood of the real space and are localizable at
scales large compared to $\ell$. The causality condition is
formulated as  continuity of the field commutator in an
appropriate topology associated with the light cone. We prove that
the relevant function spaces are nuclear and derive the kernel
theorems for the corresponding classes of multilinear functionals,
which provides the basis for the reconstruction procedure. Special
attention is given to the accurate determination of the domain of
the reconstructed quantum fields in the Hilbert space of states.
We show that the primitive common invariant domain must be
suitably extended  to  implement the (quasi)localizability and
causality conditions.

\end{abstract}

\bigskip

MSC-2000 classes:  81T05, 81T10, 46N50, 46F15, 32C81.

\smallskip

FIAN-TD/2010-03
\bigskip

\newpage

\setcounter{page}{2}

\section{Introduction}

The reconstruction theorem occupies a central position in the
Wightman axiomatic approach~\cite{SW} to quantum field theory
(QFT). This theorem establishes the conditions under which a
collection of tempered distributions is the set of vacuum
expectation values of some local field theory. Moreover, it gives
an explicit procedure for constructing the corresponding field
operators in a Hilbert space of states. This reconstruction
procedure is also applicable to nontempered distributions (known
as ultradistributions), provided that the space of test functions
on which they act contains a dense set of functions of compact
support. In that event, the reconstructed  fields belong to the
Jaffe class~\cite{Jaffe} of strictly localizable fields, for which
the local commutativity axiom, called also microcausality, can be
formulated in the usual terms. The aim of this paper is to derive
a reconstruction theorem for a larger class of  functionals that
are defined on test functions analytic in a complex
$\ell$-neighborhood of the real space. Accordingly, the
momentum-space growth of these functionals is bounded by the
exponential  $\exp(\ell|p|)$. As shown in~\cite{FS1}, such
functionals are localizable at length scales large compared to
$\ell$ and hence  $\ell$ can be regarded as a fundamental length.
A generalization of  Wightman's approach to the nonlocal field
theories with an exponential behavior of expectation values in
momentum space was first proposed by Iofa and
Fainberg~\cite{IF1,IF2}. They replaced the microcausality
condition (for the case of a single scalar field) with the
requirement of symmetry of the analytic Wightman functions under
the permutations of their arguments. Subsequently it was
shown~\cite{FS2} that a natural way of formulating causality in
nonlocal field theories with analytic test functions is by using a
suitably adapted notion of carrier of an analytic functional. This
notion plays a large part in various questions of complex and
functional analysis and is basic for the Sato-Martineau theory of
hyperfunctions, see, e.g., H\"ormander's treatise~\cite{H1}.

Although the key ideas of  nonlocal QFT with the exponential
high-energy behavior of expectation values are contained
in~\cite{IF1,IF2,FS2}, no consideration  has been given there to
the features of reconstruction of nonlocal fields. We find it
useful to close this gap because there is an intriguing intimate
connection between nonlocal field theories of this kind, string
theory~\cite{Kap} holographic models~\cite{G}, and  field theories
on noncommutative spacetime~\cite{S07, FS}. In deriving the
nonlocal version of  reconstruction theorem, we will take as
starting point the up-to-date  formulation~\cite{I} of quantum
field theory with a fundamental length in terms of vacuum
expectation values. A different strategy for reconstructing
nonlocal fields was proposed by Br\"uning and Nagamachi~\cite{BN}.
While some observations made in~\cite{BN} are very valuable, the
definition of extended local commutativity  introduced in that
paper is misleading. This definition modifies the definition of
quasilocality given in~\cite{FS2} and was motivated by the need to
adapt the latter to the nonlocal model $:e^{g\phi^2}\!\!:(x)$,
where $\phi$ is a free neutral scalar field. However a closer
examination~\cite{I} shows that the analyticity domain of the
$n$-point  functions of this model is considerably larger than
that found in~\cite{BN} and, as a consequence, the nonlocal field
$:e^{g\phi^2}\!\!:(x)$ fits in the original
framework~\cite{IF1,IF2,FS2}. It is possible even to include a
wider class of normal ordered functions of the free field, see
Sec.~VII below. Furthermore, the modified definition uses a
projective limit as $\ell\to\infty$, which makes doubtful that the
theory~\cite{BN} has the local limit when the fundamental length
approaches zero.

In the conventional formalism of local QFT, the basic test
function space is taken to be the  Schwartz space $S$ of
infinitely differentiable functions of fast decrease. An important
point is that the space $S$ is nuclear. This property is used in
deriving  most, if not all, of the results of the axiomatic
approach~\cite{SW,J,BLOT} and is  crucial for constructing the
completed tensor algebra over the  space $S(\oR^4)$, which is an
essential  ingredient of Wightman's reconstruction theorem. For
this reason, we begin the derivation of the nonlocal analog of
this theorem  with proving that the function spaces~\cite{FS2,I}
which replace $S$ in QFT with a fundamental length are nuclear.
Notice that our method of proving  nuclearity is quite general and
is also applicable to the function spaces~\cite{S07} used in field
theory on noncommutative space-time. The central problem with
reconstructing nonlocal fields is finding the exact formulation of
causality in terms of the fields, which must be equivalent to the
quasilocality condition~\cite{FS2,I} stated in terms of the
Wightman generalized functions.  It turns out that solving this
problem requires an appropriate  extension of the primitive common
invariant domain of the reconstructed field operators in the
Hilbert space.

The paper is organized as follows. In Sec.~II, the basic
definitions are given and the main result is stated. In Sec.~III,
we prove that the function spaces introduced in~\cite{FS2}
and~\cite{I} are nuclear. The kernel theorems for the
corresponding classes of multilinear functionals are established
in Sec.~IV. Making use of these results, we prove in Sec.~V the
reconstruction theorem for QFT with a fundamental length.
Particular attention is given to the precise formulation of the
quasilocality condition for the reconstructed fields, which
substitutes for the microcausality axiom of local QFT. In Sec.~VI,
we show the equivalence of this condition to that given
in~\cite{FS2,I} in terms of the vacuum expectation values.
Section~VII contains concluding remarks. Appendices A and B
present the proofs of some auxiliary statements about topological
tensor products  of locally convex vector spaces.

\section{Basic definitions and the main result}

A tempered distribution $u$ on $\oR^d$ is said to be supported in
a closed set $M\subset \oR^d$ if $(u,f)=0$ for all test functions
$f\in S(\oR^d)$ that vanish  in a neighborhood of $M$. Clearly,
this definition is inapplicable in the case of analytic test
functions. To obtain a suitable generalization, it is natural to
consider instead the property that  $(u,f)$ tends to zero as $f$
approaches zero in a neighborhood of $M$. But to formalize  this
simple idea, we must assign a topology to every open subset of
$\oR^d$, or in other words, define a presheaf of topological
vector spaces. We recall the corresponding definitions introduced
in~\cite{FS1,FS2} for nonlocal QFT with a fundamental length.

Let $\ell$ be a positive number or  $+\infty$.  For each set
$O\subset\oR^d$, we define a space  $A_\ell(O)$ in the following
way. Let $\tilde O^l$ be the complex $l$-neighborhood of $O$,
consisting of those points $z\in \oC^d$ for which there is  $x\in
O$ such that $|z-x|\equiv\max_{1\leq j\leq d}|z_j-x_j|<l$. The
space $A_\ell(O)$ consists of all  analytic functions $f$ on
${\tilde O}^\ell$ with the property that
\begin{equation}
\|f\|_{O, l, N}\defeq\max_{|\kappa|\leq N}\sup_{z\in \tilde O^l}
|z^\kappa f(z)|<\infty\qquad \text{for all}\quad
l<\ell\quad\text{and for any integer}\quad
   N.
\label{2.1}
     \end{equation}
The system of norms~\eqref{2.1} is equivalent to the countable
system $\|f\|_{O, \ell-1/N, N}$ and hence $A_\ell(O)$ is a
metrizable locally convex space. We note that the same space
corresponds to the closure ${\Bar O}$ of $O$. Clearly, every
element of $A_\ell(O)$ is completely determined by its values at
real points. The space $A_\ell(\oR^d)$ is naturally embedded in
each of $A_\ell(O)$, $O\subset \oR^d$, via the restriction map.
Let $A_\ell'(\oR^d)$ denote the topological dual of
$A_\ell(\oR^d)$, i.e., the space of all continuous linear
functionals defined on $A_\ell(\oR^d)$. According to what has been
said, a closed set $M\subset \oR^d$ can be regarded as a {\it
carrier} of $u\in A_\ell'(\oR^d)$, if $u$ has a continuous
extension to $A_\ell(M)$. By the Hahn-Banach theorem, this
property is equivalent to the continuity of $u$ in the topology
induced on $A_\ell(\oR^d)$ by that of $A_\ell(M)$. We refer the
reader to~\cite{FS1,FS2} for  a more detailed motivation of this
definition and for an  explanation why the elements of
$A_\ell'(\oR^d)$ can be thought of as localizable in $\oR^d$ at
scales large compared to $\ell$. For $O=\oR$ and
$\ell=1/(eB)<\infty$, the space $A_\ell(O)$ coincides with the
space $S^{1,B}$ in the notation used by Gelfand and
Shilov~\cite{GS2}. As shown in the  Secs.~III and IV, the
topological properties of the spaces $A_\ell(O)$ are similar to
those of the Schwartz space $S$, which makes them convenient for
use. In particular, these spaces are nuclear. Of prime importance
is the property expressed by
\begin{equation}
A_\ell(O_1)\hotimes A_\ell(O_2)=A_\ell(O_1\times O_2),
\label{2.2}
\end{equation}
where the superimposed hat is used to denote the completion of the
tensor product with respect to its natural topology. We notice
that, in general, a distinction should be made between the
projective topology and the inductive topology of a tensor
product, but in this case they coincide because any $A_\ell(O)$ is
a Fr\'echet space. It follows from~\eqref{2.2} that every
separately continuous bilinear functional on $A_\ell(O_1)\times
A_\ell(O_2)$ can be identified with a linear functional (i.e.,
with a generalized function) on $A_\ell(O_1\times O_2)$. This
statement is an analog of the famous Schwartz kernel theorem
(called sometimes nuclear theorem) for $S'$. We shall also
consider spaces with different indices and use the more general
formula
$$
A_{\ell_1}(O_1)\hotimes
A_{\ell_2}(O_2)=A_{\ell_1,\ell_2}(O_1\times O_2),
$$
where the space on the right-hand side consists of all analytic
functions on  ${\tilde O}_1^{\ell_1}\times {\tilde O}_2^{\ell_2}$
such that  the norms
\begin{equation} \|f\|_{O_1\times O_2, l_1,
l_2,N}=\max_{|\kappa|\leq N}\sup_{z\in {\tilde O}_1^{l_1}\times
{\tilde O}_2^{l_2}}
 |z^\kappa f(z)|,\quad l_1<\ell_1,\quad l_2<\ell_2,\quad  N=0,1,2,\dots,
\notag
\end{equation}
are finite.   It should be noted that the spaces $A_\ell(O)$ with
$\ell<\infty$ are not invariant under the linear coordinate
transformations and $A_\infty(\oR^d)$ is their maximal invariant
subspace.

We content ourselves with  proving a reconstruction theorem for
the case of a scalar hermitian  nonlocal field and proceed from
the formulation~\cite{I} of nonlocal QFT in terms of the Wightman
generalized functions. Before we begin, a few words about
notation. As usual, we denote by ${\mathcal P}^\uparrow_+$  the
proper orthochronous Poincar\'e group and by $f_{(a,\Lambda)}$ the
function obtained by applying a transformation $(a,\Lambda)\in
{\mathcal P}^\uparrow_+$ to $f\in A_\infty(\oR^{4n})$,
 $$
f_{(a,\Lambda)}(x_1,\dots,x_n)=
f(\Lambda^{-1}(x_1-a),\dots,\Lambda^{-1}(x_n-a)).
$$
The open  cone $\{x\in \oR^4\colon x^2=(x^0)^2- (\mathbf x)^2>0\}$
is denoted by $\oV$  and  its upper (or forward) component is
denoted by $\oV^+$. We also use  a "hat" notation to denote the
Fourier transforms of functions and functionals. Our starting
point is a set of analytic functionals $\{{\mathcal W}_n\}$
satisfying the following conditions.

\begin{enumerate}
\item[$(a.1)$] {\it Initial functional domain.}
$$
{\mathcal W}_n\in A'_\infty(\oR^{4n})\qquad \mbox{for}\quad n\geq
1.
$$

\item[$(a.2)$] {\it Hermiticity.}
$$
\overline{({\mathcal W}_n, f)}= ({\mathcal W}_n, f^\dagger)\quad
\mbox{for each}\,\, f\in A_\infty(\oR^{4n}),\quad \mbox{with}\,\,
f^\dagger(z_1,\dots,z_n)=\overline{f(\bar z_n,\dots,\bar z_1)}.
$$

\item[$(a.3)$] {\it Positive definiteness.}
\begin{equation}
\sum_{k,m=0}^N ({\mathcal W}_{k+m},f^\dagger_k\otimes f_m)\geq 0,
 \notag
 \end{equation}
where ${\mathcal W}_0=1$, $f_0\in \oC$, and $\{f_1,\dots, f_N\}$
is  an arbitrary finite set of test functions such that $f_k\in
A_\infty(\oR^{4k})$, $k=1,\dots, N$.

\item[$(a.4)$] {\it Poincar\'e covariance.}
$$
({\mathcal W}_n, f)=({\mathcal W}_n, f_{(a,\Lambda)})\qquad
\mbox{for all}\quad f\in A_\infty(\oR^{4n})\quad  \mbox{and} \quad
\mbox{for each} \quad
 (a,\Lambda)\in {\mathcal P}^\uparrow_+.
$$
This property is equivalent to the existence of Lorentz invariant
functionals $W_n\in A'_\infty(\oR^{4(n-1)})$ such that
\begin{equation}
{\mathcal W}_n(x_1,\dots x_n)=W_n(x_1-x_2,\dots,x_n-x_{n-1}),
\quad n\geq1.
 \notag
\end{equation}

\item[$(a.5)$] {\it Spectral condition.}
$$
\supp \hat{W}_n\subset
\underbrace{\bar{\oV}^+\times\dots\times\bar{\oV}^+}_{(n-1)}.
$$

\item[$(a.6)$] {\it Cluster decomposition property.\/} If $a$ is a
spacelike vector, then for each $f\in A_\infty(\oR^{4k})$ and for
each $g\in A_\infty(\oR^{4m})$,
$$
({\mathcal W}_{k+m},f\otimes g_{(\lambda a,I)})\longrightarrow
({\mathcal W}_k,f)({\mathcal W}_m,g) \qquad \mbox{as}\quad
\lambda\to \infty.
$$

\item[$(a.7.1)$] {\it Quasilocalizability.\/} There exists
$\ell<\infty$ such that every functional $W_n$ has a continuous
extension to the space $A_\ell(\oR^{4(n-1)})$.

  \item[$(a.7.2)$] {\it Quasilocality.\/}
  For any $n\geq2$ and $1\leq k\leq n-1$, the difference
\begin{multline}
W_n(\xi_1,\dots,\xi_{k-1},\xi_k,\xi_{k+1},\dots, \xi_{n-1})\\ -
W_n(\xi_1,\dots,\xi_{k-1}+\xi_k,-\xi_k,\xi_k+\xi_{k+1},\dots,
\xi_{n-1})
 \label{2.3}
 \end{multline}
extends continuously to the space $A_\ell(V_{(k)})$, where
\begin{equation}
V_{(k)}=\{\xi\in\oR^{4(n-1)}\colon \xi_k^2> 0\}. \label{2.4}
\end{equation}

\end{enumerate}

It is significant that conditions (a.7.1) and (a.7.2) are
formulated with respect to the relative coordinates
$\xi_k=x_k-x_{k+1}$, $k=1,\dots n-1$, see~\cite{I} for a
discussion of this point. The reconstruction theorem will be
divided into three parts represented by Theorems~1--3, whose
proofs are given in Sec.~V.

{\bf Theorem~1:} {\it  Let $\{{\mathcal W}_n\}$, $n=1,2,\dots$, be
a sequence of analytic functionals satisfying  conditions
(a.1)--(a.6). Then there exist a separable Hilbert space $\mathcal
H$, a continuous unitary representation  $U(a,\Lambda)$ of the
group ${\mathcal P}^\uparrow_+$ in $\mathcal H$, a unique state
$\Psi_0$ invariant under $U(a,\Lambda)$, and a hermitian scalar
field $\varphi$ with an invariant  dense domain $D\subset \mathcal
H$ such that
\begin{equation}
\langle \Psi_0,\, \varphi(f_1)\ldots \varphi(f_n)\Psi_0\rangle =
 \notag (\mathcal W_n, f_1\otimes\dots\otimes f_n),\qquad
 f_j\in A_\infty(\oR^4),\quad j=1,\dots,n.
\end{equation}
The field  $\varphi$ obeys  all Wightman's axioms except for
locality and with $A_\infty(\oR^4)$ instead of the Schwartz space.
Any other relativistic quantum field theory with the same vacuum
expectation values is unitary equivalent to this one.}

Since we have an analog of Schwartz's kernel theorem, the proof of
Theorem~1 is completely analogous to that of the corresponding
part of the Wightman reconstruction theorem~\cite{SW}.

{\bf Theorem~2:} {\it From condition (a.7.1) it follows that the
operator-valued generalized function $\varphi(f)$ defined on $D$
by Theorem~1 with  $f\in A_\infty(\oR^4)$ extends continuously to
the space  $A_{\ell/2}(\oR^4)$. Moreover, this condition is
equivalent to the fact that every  monomial $\varphi(f_1)\dots
\varphi(f_n)$, $n\geq1$, can   be uniquely extended to an
operator-valued generalized function over the space
$A^{(r)}_\ell(\oR^{4n})$ consisting of all functions  of the form
}
\begin{equation}
g^{(r)}(x_1,\dots,x_n)= g(x_1, x_1-x_2,\dots,x_{n-1}-x_n),\quad
\text{where $g\in A_{\ell/2,\ell}(\oR^4\times \oR^{4(n-1)})$}.
\label{2.5}
\end{equation}

(The index $r$ indicates changing to the relative coordinates.)

{\it Remark~1:} The extension is unique because
$A_\infty(\oR^{4n})$ is dense in
$A_{\ell/2,\ell}(\oR^4\times\oR^{4(n-1)})$. A simple proof of the
fact that $A_\infty(\oR^d)$ is dense in $A_\ell(\oR^d)$ for any
$\ell$ and $d$ is given in Appendix of~\cite{I}, and after minor
changes in the notation, this proof applies also to the spaces
$A_{\ell_1, \ell_2}(\oR^{d_1}\times \oR^{d_2})$, $d_1+d_2=d$.
Besides the space $A_\infty(\oR^{4n})$ is invariant under linear
transformations of $\oR^{4n}$, therefore it is dense in
$A_\ell^{(r)}(\oR^{4n})$.

{\it Corollary: The operator-valued generalized function
$\varphi(f)$  can be uniquely extended, with preserving
hermiticity and the continuity in $f\in A_{\ell/2}(\oR^4)$, to the
domain $D_\ell$ spanned by $D$ and all vectors of the form
\begin{equation}
\Psi(g^{(r)})=\int {\rm}dx_1\dots {\rm}dx_n\,
g^{(r)}(x_1,\dots,x_n)\prod_{i=1}^n\varphi(x_i)\Psi, \notag
\end{equation}
where $g\in A_{3\ell/2,\ell}(\oR^4\times \oR^{4(n-1)})$ and
$\Psi\in D$.

 Proof:} If  $f\in A_{\ell/2}(\oR^4)$  and $g\in
A_{3\ell/2,\ell}(\oR^4\times\oR^{4(n-1)})$, then the function
\begin{equation}
h(x,x_1,\dots,x_n)= f(x)g(x-x_1, x_2,\dots, x_n) \notag
\end{equation}
belongs to $ A_{\ell/2,\ell}(\oR^4\times \oR^{4n})$ and hence
$h^{(r)}= f\otimes g^{(r)}$ belongs to
$A^{(r)}_\ell(\oR^{4(n+1)})$. Therefore, we can define an
extension of $\varphi(f)$ to $D_\ell$ by $\Psi(g^{(r)})\to
\Psi(f\otimes g^{(r)})$ with subsequent  extension by linearity.
However, we must show that $\Psi(g^{(r)})=0$ implies
$\Psi(f\otimes g^{(r)})=0$. To this end, we approximate $f$ by a
sequence $f_\nu\in A_\infty(\oR^4)$ and $g$ by a sequence
$g_\nu\in A_\infty(\oR^{4n})$. Let $\Phi$ be an arbitrary element
of $D$. Using  the hermiticity of $\varphi(f_\nu)$ and Theorem~2,
we obtain
\begin{equation}
\langle\Phi, \Psi(f_\nu\otimes
g^{(r)}_\nu)\rangle=\langle\varphi(f_\nu^\dagger)\Phi,
\Psi(g^{(r)}_\nu)\rangle\to\langle\varphi(f^\dagger)\Phi,
\Psi(g^{(r)})\rangle=0.
 \notag
\end{equation}
So, the required implication holds and we conclude that the
extension of $\varphi(f)$ is well defined  and continuous in $f\in
A_{\ell/2}(\oR^4)$. A similar reasoning with an additional
approximation $\Phi_\nu\to\Phi\in D_\ell$ shows that the extension
(which we denote by the same symbol) satisfies $\langle
\Phi,\varphi(f)\Psi\rangle=
\langle\varphi(f^\dagger)\Phi,\Psi\rangle$ for all $\Phi,\Psi\in
D_\ell$.  Now let $\psi(f)$ be another  extension of $\varphi(f)$
to $D_\ell$, which is continuous in the topology of
$A_{\ell/2}(\oR^4)$. Using again the approximating sequences
$f_\nu$ and $g_\nu$ and the  hermiticity of $\varphi(f_\nu)$, we
infer that   $\langle
\Phi,(\psi(f)-\varphi(f))\Psi(g^{(r)})\rangle=0$ for all $\Phi\in
D$,  for every $g\in A_{3\ell/2,\ell}(\oR^4\times \oR^{4(n-1)})$,
and for each $n=1,2,\dots$. This completes the proof of Corollary
of Theorem~2.

 It should be noted that $D_\ell$ is not  invariant either
under the action of operators $\varphi(f)$, $f\in
A_{\ell/2}(\oR^4)$, or under the  Lorentz group, but it follows
from $(a.4)$ and $(a.7.1)$ that the operator-valued generalized
function $\varphi(f)$ extends continuously to each space
obtainable from $A_{\ell/2}(\oR^4)$ by a Lorentz transformation
$\Lambda$ and this extension is well defined on
$U(0,\Lambda)D_\ell$.

{\bf Theorem~3:} {\it Let $\varphi$ be the quantum field
constructed by Theorem~1 and let $D_\ell$ be the domain specified
in Corollary of Theorem~2. From condition  (a.7.2) it follows that
for any $\Psi,\Phi\in D_\ell$, the bilinear functional
\begin{equation}
\langle\Phi,[\varphi(f_1), \varphi(f_2)]_-\Psi\rangle
\label{2.6}
\end{equation}
on $A_{\ell/2}(\oR^4)\times A_{\ell/2}(\oR^4)$, which
by~\eqref{2.2} is identified with an element of
$A'_{\ell/2}(\oR^{4\cdot2})$, extends continuously to the space
$A_{\ell/2}(\oW)$, where}
\begin{equation}\oW=\{(x,x')\in \oR^{4\cdot2}\colon
(x-x')\in \oV \}. \label{2.7}
\end{equation}

\section{Nuclearity of the spaces  $A_\ell(O)$}

To prove Theorems~1--3, we need the following result.

{\bf Theorem~4:} {\it  $A_\ell(O)$ is a nuclear Fr\'echet space
for each $\ell$ and for any  $O\subset \oR^d$.

Proof:} Let  $A_{l, N}(O)$ be the space  of all analytic functions
$f$ on $\tilde O^l$ with the property that $\|f\|_{O, l,
N}<\infty$. It is easily seen that this normed space is complete.
Indeed, if $f_\nu $ is a Cauchy sequence in $A_{l, N}(O)$, then it
converges uniformly on $\tilde O^l$ and hence the limit function
$f(z)=\lim_{\nu\to\infty} f_\nu(z)$ is analytic in this domain.
There is a constant $C$ such that $\|f_\nu\|_{O, l, N}\leq C$.
Therefore, $|z^\kappa f(z)|\leq C$ for $|\kappa|\leq N$ and for
all $z\in\tilde O^l$. So, $f\in A_{l, N}(O)$. The space
$A_{\ell}(O)$, being the projective limit of the complete spaces
$A_{l, N}(O)$, is also complete. Besides, as noted above, it is
metrizable and is hence a Fr\'echet space. It remains to show that
$A_\ell(O)$ is nuclear. To this end, we recall some facts from
functional analysis.

A  convenient criterion of nuclearity is formulated by
Pietsch~\cite{P} in terms of the Radon measure defined on the
polars of neighborhoods of the origin. Let $F$ be a locally convex
space, $F'$ be its dual, and  $V\subset F$. The set of functionals
$u\in F'$ such that $\sup_{f\in V}|(u,f)|\leq 1$ is called the
(absolute) polar of $V$ and denoted by $V^\circ$. The polar of
each neighborhood of 0 is compact under the weak topology
$\sigma(F',F)$, see Schaefer's textbook~\cite{Sch}. A Radon
measure on a compact set $Q$ is, by definition, a continuous
linear form on the space $C(Q)$ of continuous functions on $Q$. A
Radon measure $\mu$ is called positive if $\mu(\psi)\geq 0$ for
all non-negative functions $\psi\in C(Q)$. Let $f\in F$ and let
$p_f$ be the semi-norm on $F'$ defined by $p_f(u)=|(u,f)|$.  The
function $p_f(u)$ is continuous in the topology $\sigma(F',F)$ by
the definition of the latter. If $U$ is an absolutely convex
absorbing set in $F$, then its associated  Minkowski functional
$p_U$ is defined by
\begin{equation} p_U(f)=\inf\{t>0\colon f\in tU\}.
 \label{3.1}
\end{equation}
By the Pietsch theorem, a locally convex space $F$ is nuclear if
and only if  for every absolutely convex neighborhood $U$ of 0 in
$F$, there is an absolutely convex neighborhood $V$ of 0 and a
positive Radon measure $\mu$ on $V^\circ$ such that, for all $f\in
F$,
\begin{equation}
p_U(f)\le  \mu(p_f|_{V^\circ}),
 \label{3.2}
\end{equation}
where $p_f|_{V^\circ}$  is the restriction of $p_f$ to $V^\circ$.
[In~\cite{P}, Sec.~4.1.5, the right-hand side of~\eqref{3.2} is
written as $\int_{V^\circ}|(u,f)|\,d\mu$.]

In order to make use of this theorem, we represent the topology of
$A_\ell(O)$ in a  different form. Namely,  the system of
norms~\eqref{2.1} is equivalent to the system
\begin{equation}
\|f\|'_{O, l, N}=\int_{\tilde O^l}
 (1+|z|)^N |f(z)|\,{\rm d}x{\rm d}y,\qquad  l<\ell,\quad
 N=0,1,2,\dots\quad (z=x+iy).
 \label{3.3}
\end{equation}
Indeed, taking into account that
\begin{equation}
(1+|z|)^N\leq 2^N\max(1, |z|^N)=2^N \max_{\kappa\leq N}|z^\kappa|,
 \label{3.4}
\end{equation}
we obtain
\begin{equation}
\|f\|'_{O, l, N}\leq \left(\int_{|y|\leq l}\frac{{\rm d}x {\rm
d}y}{(1+|z|)^{d+1}}\right)\sup_{z\in {\tilde O}^l}(1+|z|)^{N+d+1}
|f(z)|\leq C\,\|f\|_{O, l, N+d+1}.
 \notag
\end{equation}
On the other hand, the function $z^\kappa f(z)$ is analytic on
${\tilde O}^\ell$ and we may use Theorem~2.2.3 in~\cite{H2}, which
shows that for each $l'$ satisfying $l<l'<\ell$, there is a
constant $C'$ such that
\begin{multline}
\|f\|_{O, l, N}\leq C'\max_{\kappa\leq N}\sup_{\zeta\in {\tilde
O}^l}\int_{|z-\zeta|\leq l'-l}|z^\kappa f(z)|\,{\rm d}x{\rm d}y\\
\leq C'\int_{{\tilde O}^{l'}}(1+|z|)^N|f(z)|\,{\rm d}x{\rm d}y=
C'\,\|f\|'_{O, l', N}.
 \notag
\end{multline}
Thus, every absolutely convex neighborhood $U$ of 0 in $A_\ell(O)$
contains a neighborhood of the form $\|f\|'_{O,l,N}<\epsilon$ and
hence its Minkowski functional $p_U$ satisfies the inequality
\begin{equation}
p_U(f)\leq \epsilon^{-1}\|f\|'_{O,l,N}.
 \label{3.5}
\end{equation}
Now we apply the Pietsch theorem, taking
\begin{equation}
V=\{f\in A_\ell(O)\colon \sup_{{\tilde O}^l}(1+|z|)^{N+d+1}
|f(z)|<1\}. \label{3.6}
\end{equation}
Using~\eqref{3.4}, we see that $V$ contains all functions $f$ such
that $\|f\|_{O,l,N+d+1}<2^{-(N+d+1)}$ and is hence  a neighborhood
of 0. Let $z\in {\tilde O}^l$ and let $\delta_{z,N}$ be the
continuous linear form on $A_\ell(O)$ defined by
$$
(\delta_{z,N},f)=(1+|z|)^{N+d+1}f(z).
$$
Clearly,  $\delta_{z,N}\in V^\circ$  and the map ${\tilde O}^l\to
A'_\ell(O)\colon z\to \delta_{z,N}$ is continuous in the topology
$\sigma(A'_\ell(O),A_\ell(O))$. If $\psi$ is a continuous function
on $V^\circ$, then the function $z\to\psi(\delta_{z,N})$ is
continuous and bounded on ${\tilde O}^l$ for any $N$. This enables
us to define a Radon measure $\mu$ on $C(V^\circ)$ by the formula
\begin{equation}
\mu(\psi)=\epsilon^{-1}\int_{{\tilde O}^l} \frac{\psi(\delta_{z,
N})}{(1+|z|)^{d+1}}\,{\rm d}x {\rm d}y.
 \label{3.7}
   \end{equation}
Functional~\eqref{3.7} is obviously bounded and positive.
Furthermore, we have
\begin{equation}
 \mu(p_f)=\epsilon^{-1}\int_{{\tilde O}^l}
(1+|z|)^N|f(z)\,|{\rm d}x {\rm d}y= \epsilon^{-1}\|f\|'_{O,l,N}.
 \label{3.8}
   \end{equation}
It follows from~\eqref{3.5} and \eqref{3.8} that
condition~\eqref{3.2} is satisfied with this choice of  $V$ and
$\mu$. This completes the proof of Theorem~4.

{\it Corollary: For any $O\subset \oR^d$, the space $A_\ell(O)$ is
a Fr\'echet-Schwartz ($FS$) and hence Montel  space. In
particular, it is barrelled, reflexive and separable.}

We recall~\cite{Sch} that any nuclear Fr\'echet space can be
represented as the projective limit of a decreasing sequence of
Hilbert spaces with nuclear connecting maps. Every nuclear map is
compact and the projective limit of a decreasing sequence of
locally convex spaces with compact connecting maps is an FS space.
These two classes of spaces have been introduced by
Grothendieck~\cite{Grot2,Grot1} A description of the properties of
FS spaces is given, e.g., in Morimoto's monograph~\cite{Mor}.

{\it Remark~2:} Clearly, $A_{\ell_1,\ell_2}(O_1\times O_2)$ is
also a nuclear Fr\'echet space for any $\ell_1,\ell_2$, and $O_1,
O_2$. The proof is the same as above, with obvious changes in
notation.

{\it Remark~3:} Theorem~4 gives a  new  simple proof of the
well-known fact~\cite{GS4} that the spaces $S^{1,
B}(\oR^d)=A_{1/eB}(\oR^d)$ are nuclear.  In particular, so is
$S^{1, 0}(\oR^d)=A_\infty(\oR^d)$, which is the test function
space for the ultra-hyperfunctions used in~\cite{BN}. We also note
that the nuclearity of $S^{1, B}(\oR^d)$ implies the nuclearity of
$S^1(\oR^d)=\injlim_{B\to\infty} S^{1,B}(\oR^d)$ by the hereditary
properties of the inductive limits of countable families of
locally convex spaces. This simple proof of nuclearity applies
also to all Gelfand-Shilov spaces $S^\beta$ and $S^\beta_\alpha$
with  $\beta<1$ and  is easily adaptable to the spaces with
$\beta>1$.

The fact that  every $A_\ell(O)$ is an  FS space can  be used in
deriving other important properties of this presheaf of spaces. As
an example, we prove the following decomposition theorem.

{\bf Theorem~5:} {\it Let  $O_1$ and  $O_2$ be any two sets in
$\oR^d$. If a functional $u\in A'_\ell(\oR^d)$ has a continuous
extension to  $A_\ell(O_1\cup O_2)$, then it can be decomposed as
$u=u_1+u_2$, where $u_1$ and $u_2$ extend,  respectively, to
$A_\ell(O_1)$ and  $A_\ell(O_2)$.

Proof:} We  consider  $A_\ell(O_1\cup O_2)$ as a linear subspace
of  $A_\ell(O_1)\times A_\ell(O_2)$, by assigning to each  $f\in
A_\ell(O_1\cup O_2)$ the pair of restrictions $f|\tilde O_1^\ell$,
$f|\tilde O_2^\ell$. This subspace is closed because  coincides
either with the whole product space or with the kernel of the
continuous map that takes each pair $(f_1,f_2)\in
A_\ell(O_1)\times A_\ell(O_2)$ to the difference $f_1-f_2$
belonging to the space of all rapidly decreasing analytic
functions on $\tilde O_1^\ell\cap \tilde O_2^\ell$. As known, the
product of a finite number of Fr\'echet spaces is a Fr\'echet
space and so is any closed  subspace of a Fr\'echet space. It
follows that the topology induced on $A_\ell(O_1\cup O_2)$ by that
of $A_\ell(O_1)\times A_\ell(O_2)$ coincides with its original
topology by the open mapping theorem. By the Hahn-Banach theorem
the functional $u$ has a continuous extension $\tilde u$ to the
product space. Therefore we can write  $u(f)= \tilde u(f|\tilde
O_1^\ell,0)+ \tilde u(0,f|\tilde O_2^\ell)$, which completes the
proof because the injections $A_\ell(O_{1,2})\to A_\ell(O_1)\times
A_\ell(O_2)$ are continuous.

\section{Kernel theorem for the spaces  $A_\ell(O)$}

Now we  recall some basic facts~\cite{Sch} about the tensor
products of locally convex spaces. Let $F$ and  $G$ be such
spaces. By the definition of the tensor product $F\otimes G$,
there is a canonical bilinear map $(f,g)\to f\otimes g$ from
$F\times G$ to $F\otimes G$, which is continuous if $F\otimes G$
is equipped with the projective topology $\tau_\pi$. Furthermore,
$F\otimes_\pi G$ has the following universality property: For any
locally convex space $E$ and for each continuous bilinear map
$\beta\colon F\times G\to E$, there is a unique continuous linear
map $\beta_*\colon F\otimes_\pi G\to E$ such that
$\beta_*(f\otimes g)=\beta(f,g)$ for all $f\in F$ and $g\in G$.
The linear map $\beta_*$ is called the map associated with
$\beta$. The completion of $F\otimes_\pi G$ is denoted by
$F\mathbin{\hat{\otimes}}G$.

The next theorem develops Grothendieck's construction given
in~\cite{Grot2}, Chap.~2, Th\'eoreme~13.

{\bf Theorem~6:} {\it  Let $F$, $G$, and $H$ be complete locally
convex spaces  of scalar functions defined, respectively, on $X$,
$Y$, and $X\times Y$.  Let the topology of each of these spaces be
not weaker than the topology of pointwise convergence. Suppose G
is nuclear,  $H$ is barrelled, and the following conditions are
satisfied:
\begin{itemize}
 \item[$(i)$] For any $f\in F$ and $g\in G$, the function $(x,y)\to
f(x)\,g(y)$ belongs to $H$ and the corresponding bilinear map
$\omega\colon F\times G\to H$ is continuous;
 \item[$(ii)$] For any
$h\in H$ and for each $x\in X$, the function  $y\to h(x,y)$
belongs to $G$ and, if $v\in G'$, the function $h_v \colon x \to
(v, h(x,\cdot))$ belongs to $F$;
 \item[$(iii)$] The bilinear map
$G'\times H\to F\colon (v,h)\to h_v$ is separately continuous if
$G'$ is equipped with the strong topology.
\end{itemize}
Then $F\mathbin{\hat{\otimes}}G$ can be identified with  $H$.}

The proof of this theorem is presented in Appendix A. We shall
soon see that the above conditions are easily verified  and this
gives a simple derivation of the desired kernel theorem for the
class of spaces we are working with. It is worth noting that if
$F$, $G$, and $H$ are Fr\'echet spaces, then $(iii)$ is a
consequence of $(i)$ and $(ii)$, see~\cite{SS2} or~\cite{Smirn}.

{\bf Theorem~7:} {\it For any $O_1\subset \oR^{d_1}$ and
$O_2\subset\oR^{d_2}$, the space
$A_{\ell_1}(O_1)\mathbin{\hat{\otimes}}A_{\ell_2}(O_2)$ is
identified with $A_{\ell_1,\ell_2}(O_1\times O_2)$.

Proof.} We assume for simplicity that $\ell_1=\ell_2=\ell$ because
the proof for $\ell_1\ne\ell_2$  is in essence the same but
cumbersome in notation. We use Theorem~6 with $F= A_\ell(O_1)$,
$G= A_\ell(O_2)$, and $H=A_\ell(O_1\times O_2)$. All these spaces
are complete and nuclear by Theorem~4. We also recall that every
Fr\'echet space is barrelled. Condition $(i)$ is obviously
fulfilled because $(\widetilde{O_1\times O_2})^l=\tilde O^l\times
\tilde O^l$ and we have the inequality
\begin{equation}
\|f(z_1)g(z_2)\|_{O_1\times O_2,
l,N}\leq\|f\|_{O_1,l,N}\|g\|_{O_2,l,N},
 \label{4.1}
   \end{equation}
which demonstrates that the bilinear map $\omega\colon (f, g)\to
f\otimes g$ is continuous at  $(0,0)$ and hence everywhere. Now
let $h\in A_\ell(O_1\times O_2)$ and $v\in A'_\ell(O_2)$. Then
there are $l_2$ and $N_2$ such that
\begin{equation}
\|v\|_{O_2,l_2,N_2}\defeq\sup_{f\in
A_\ell(O_2)}\frac{|(v,f)|}{\,\,\|f\|_{O_2,l_2,N_2}}<\infty.
 \label{4.2}
\end{equation}
 The function $h_v(z_1)=(v,h(z_1,\cdot))$ satisfies
\begin{equation}
\|h_v\|_{O_1, l_1, N_1}=\max_{|\kappa|\leq N_1}\sup_{z_1\in \tilde
O^{l_1}_1}
 |z_1^\kappa h_v(z_1)|\leq
  \|v\|_{O_2,l_2,N_2}\|h\|_{O_1\times O_2,\max(l_1,l_2),N_1+N_2}
 \label{4.3}
   \end{equation}
for each $l_1<\ell$ and for all $N_1=0,1,\dots$. We must show that
$h_v(z_1)$ is analytic on $\tilde O_1^\ell$, i.e., has all partial
derivatives in the complex variables $z_{1j}$, $j=1,\dots, d_1$,
at each point of this domain. Let $z_1\in\tilde O_1^l$,
$l<l'<\ell$, and let $L$ be  the segment of the straight line
joining the points $z_{1j}$ and  $z_{1j}+\Delta z_{1j}$, where
$|\Delta z_{1j}|<(l'-l)/2$. The corresponding increment of the
function $h$ can be written as $\Delta_{z_{1j}}h(z_1,z_2)=\int_L
h'_{z_{1j}}(z_{11},\dots,\zeta_{1j},\dots z_{1d_1},
z_2)\,d\zeta_{1j}$.  Let $\tilde L^r$ be the complex
$r$-neighborhood of $L$, with $r\leq (l'-l)/2$. Using the Cauchy
inequality, we obtain
\begin{equation}
|\Delta_{z_{1j}}h(z_1,z_2)|\leq r^{-1}|\Delta
z_{1j}|\sup_{\zeta_{1j}\in{\tilde
L}^r}|h(z_{11},\dots,\zeta_{1j},\dots z_{1d_1}, z_2)|. \notag
   \end{equation}
Therefore, the difference quotient  $\Delta_{z_{1j}}h/\Delta
z_{1j}$ considered at fixed $z_1\in \tilde O_1^l$ as an element of
$A_\ell(O_2)$ (parametrically depending on $\Delta z_{1j}$)
satisfies the inequality
\begin{equation}
\left\|\frac{\Delta_{z_{1j}}h}{\Delta
z_{1j}}\right\|_{O_2,l,N}\leq \frac{2}{l'-l}\,\|h\|_{O_1\times
O_2,l',N}
 \notag
   \end{equation}
for any $N$. Thus, the set of difference quotients is bounded in
$A_\ell(O_2)$. It also follows from the Cauchy inequality   that
$h'_{z_{1j}}\in A_\ell(O_1\times O_2)$. Using that $A_\ell(O_2)$
is a Montel space, we can choose a convergent sequence from the
set of difference quotients. Clearly, its limit is $h'_{z_{1j}}$
because the topology of $A_\ell(O_2)$ is stronger than that of
pointwise convergence. The uniqueness of this limit implies that
the difference quotient converges to $h'_{z_{1j}}$ in
$A_\ell(O_2)$.  Therefore, $(v,\Delta_{z_{1j}}h/\Delta z_{1j})\to
(v,h'_{z_{1j}})$ as $\Delta z_{1j}\to 0$ and we conclude that
condition $(ii)$ is fulfilled.

Estimate~\eqref{4.3} shows that the map $h\to h_v$ is continuous
for every fixed $v$. Now we hold $h$  fixed and let $v$ belong to
the space $A'_{l_2,N_2}(O_2)$ of functionals with finite
norm~\eqref{4.2}.  It follows also from~\eqref{4.3} that the map
$A'_{l_2,N_2}(O_2)\to A_\ell(O_1)\colon v\to h_v$ is continuous
for each $l_2<\ell$ and for every integer $N_2$. This amounts to
saying that the corresponding map $A'_\ell(O_2)\to A_\ell(O_1)$ is
continuous in the inductive limit topology determined on
$A'_\ell(O_2)$ by the canonical injections $A'_{l_2,N_2}(O_2)\to
A'_\ell(O_2)$. But this topology coincides with the strong
topology of $A'_\ell(O_2)$ by   the general open mapping
theorem~\cite{K}. In fact, even Grothendieck's version (given in
Introduction of~\cite{Grot2}) of this theorem  applies here
because the strong dual of any FS space is a dual
Fr\'echet-Schwartz (DFS) space (see~\cite{Mor}) and  is hence
ultrabornological or of type $(\beta)$ in the terminology used by
Grothendieck. This completes the proof of Theorem~7.

Because the projective tensor product is associative~\cite{K} it
follows from Theorem~7 that
\begin{equation}
A_{\ell_1}(O_1)\mathbin{\hat{\otimes}}\dots
\mathbin{\hat{\otimes}} A_{\ell_n}(O_n) =
A_{\ell_1,\dots,\ell_n}(O_1\times\dots\times O_n)\label{4.4}
   \end{equation}
for any $\ell_j>0$ and $O_j\subset \oR^{d_j}$, $j=1,\dots, n$.

 {\it Corollary [Kernel theorem for the
spaces $A_\ell(O)$]:  Let $\mu$ be a separately continuous
multilinear map of $A_{\ell_1}(O_1)\times\dots\times
A_{\ell_n}(O_n)$ into a locally convex space $E$. Then there is a
unique continuous linear map $u_\mu\colon
A_{\ell_1,\dots,\ell_n}(O_1\times\dots\times O_n)\to E$ such that
\begin{equation}
\mu(f_1,\dots, f_n)=(u_\mu, f_1\otimes\dots\otimes f_n)
 \notag
\end{equation} for any $f_j\in A_{\ell_j}(O_j)$, $j=1,\dots,
n$.}

For $n=2$, this  follows immediately from Theorem~7 because every
separately continuous bilinear map of the product of two Fr\'echet
spaces into a topological vector space is continuous~\cite{Sch}.
For $n>2$, we argue by induction, using the following simple
lemma, which is a generalization of Lemma~3 in~\cite{S04}.

{\bf  Lemma~1:} {\it Let $F$, $G$, and $E$ be locally convex
spaces and let $L$ be a sequentially dense subspace of $F$.
Suppose in addition that  $G$ is barrelled and $E$ is Hausdorff
and complete. Then every  separately continuous bilinear map
$\beta\colon L\times G\to E$ has a unique extension to a
separately continuous bilinear map $F\times G\to E$.

Proof:} For each fixed  $g\in G$,  the linear map $L\to E\colon
f\to \beta(f,g)$ can be uniquely extended to $F$ by continuity and
thereby we obtain  a map $\hat \beta\colon F\times G\to E$. We
have only to show that $\hat \beta$ is linear and continuous in
$g$ for every fixed $f\in F$. We choose a sequence $f_\nu\in L$
convergent to $f$ and consider the corresponding sequence of
continuous linear maps $G\to E\colon g\to \beta (f_\nu,g)$. This
sequence of maps converges pointwise  to the map $g\to \hat
\beta(f,g)$ and hence the latter is linear and continuous by the
Banach-Steinhaus theorem because $G$ is barrelled. The lemma is
thus proved.

To complete the proof of the kernel theorem for $A_\ell(O)$, we
use Lemma~1 with $F=A_{\ell_1}(O_1)\mathbin{\hat{\otimes}}\dots
\mathbin{\hat{\otimes}} A_{\ell_{n-1}}(O_{n-1})$,
$G=A_{\ell_n}(O_n)$, and  $L=A_{\ell_1}(O_1)\mathbin{\otimes}\dots
\mathbin{\otimes} A_{\ell_{n-1}}(O_{n-1})$.

{\it Remark~4:} This result contains, as a simple special case
(with $\ell_1=\dots=\ell_n=\infty$, $O_1=\dots= O_n=\oR^4$, and
$E$ a Banach space)  the kernel theorem for ultra-hyperfunctions
discussed by Br\"uning and Nagamachi~\cite{BN}. Because their main
effort was to cover multilinear maps with values in a Banach
space,  it is worth noting that the general kernel theorem for
multilinear maps taking values in an arbitrary locally convex
space  is an immediate consequence of its simplest version for the
complex-valued bilinear forms, see Appendix~B.

It cannot be asserted that $A_\infty(\oR^d)$ is dense in
$A_\ell(O)$ for an arbitrary domain  $O\subset \oR^d$, although
this is the case for $O=\oR^d$. For this reason it is useful to
introduce another class of spaces. Namely, we define ${\mathcal
A}_\ell(O)$ to be the closure of $A_\infty(\oR^d)$ regarded as a
subspace of $A_\ell(O)$ and provide it with the topology induced
by that of $A_\ell(O)$.  Since $A_\infty(\oR^d)$ is dense in
$A_\ell(\oR^d)$, the space ${\mathcal A}_\ell(O)$ can also be
defined as the completion of $A_\ell(\oR^d)$ with respect to the
topology induced on it by that of $A_\ell(O)$. Analogously,
$\mathcal A_{\ell_1,\ell_2}(O_1\times O_2)$ is the closure of
$A_\infty(\oR^{d_1}\times \oR^{d_2})$ in
$A_{\ell_1,\ell_2}(O_1\times O_2)$.

{\bf Theorem~8:} {\it Every space ${\mathcal A}_\ell(O)$ is a
nuclear Fr\'echet space and, for any  $\ell_1,\ell_2>0$ and for
any $O_1\subset \oR^{d_1}$, $O_2\subset \oR^{d_2}$, there is the
canonical isomorphism
\begin{equation}
\mathcal A_{\ell_1}(O_1)\hotimes \mathcal
A_{\ell_2}(O_2)\cong\mathcal A_{\ell_1,\ell_2}(O_1\times O_2).
\label{4.5}
   \end{equation}

   Proof:} The first statement of the theorem follows from the
well-known   hereditary properties~\cite{Sch} of Fr\'echet and
nuclear spaces. Now let  $f\in\mathcal A_{\ell_1}(O_1)$,
$g\in\mathcal A_{\ell_2}(O_2)$, and let sequences $f_\nu\in
A_\infty(\oR^{d_1})$, $g_\nu\in A_\infty(\oR^{d_2})$ be such that
$f_\nu\to f$ in $A_{\ell_1}(O_1)$ and $g_\nu\to g$ in
$A_{\ell_2}(O_2)$. Using an analog of  inequality~\eqref{4.1} for
the case $\ell_1\ne\ell_2$, it is easy to see that $f_\nu\otimes
g_\nu\to f\otimes g$ in the topology of
$A_{\ell_1,\ell_2}(O_1\times O_2)$. Therefore, $f\otimes g$
belongs to $\mathcal A_{\ell_1,\ell_2}(O_1\times O_2)$ and
condition $(i)$ of Theorem~6 is satisfied. Now let $h\in \mathcal
A_{\ell_1,\ell_2}(O_1\times O_2)$, $h_\nu\in
A_\infty(\oR^{d_1}\times\oR^{d_2})$, and  $h_\nu\to h$  in $
A_{\ell_1,\ell_2}(O_1\times O_2)$. Then for every fixed $z_1$, the
sequence   $h_\nu(z_1,z_2)$ tends to the function $z_2\to
h(z_1,z_2)$ in  the topology of $A_{\ell_2}(O_2)$ and hence this
function belongs to $\mathcal A_{\ell_2}(O_2)$. Let  $v\in
\mathcal A'_{\ell_2}(O_2)$. Arguing as in the proof of Theorem~7,
we see that the functions $z_1\to (v,h_\nu(z_1,\cdot))$ belong to
$A_\infty(\oR^{d_1})$ because
$\|v\|_{\oR^{d_2},l_2,N_2}\leq\|v\|_{O_2,l_2,N_2}$. It follows
from (an analog of) \eqref{4.3} that the sequence of these
functions converges to the function $z_1\to (v,h(z_1,\cdot))$ in
the topology of $A_{\ell_1}(O_1)$  as  $\nu\to\infty$. So
condition $(ii)$ of Theorem~6 is  fulfilled. Since $\mathcal
A_{\ell_2}(O_2)$ is an FS space, we can  now  apply the arguments
used at the end of the proof of Theorem~7 and conclude that
condition $(iii)$ of Theorem~6 is also fulfilled. This completes
the proof.

\section{Proof of the reconstruction theorem}

The kernel theorem enables us to construct a tensor algebra
$T(A_\infty(\oR^4))$ in complete analogy with constructing the
Borchers algebra associated  with the Schwartz space $S(\oR^4)$.
Namely,
\begin{equation}
T(A_\infty(\oR^4))= \bigoplus_{n=0}^\infty T_n, \quad
T_0=\oC,\quad
T_n=A_\infty(\oR^4)^{\hat{\otimes}n}=A_\infty(\oR^{4n}) \quad
\mbox{for}\quad n\geq 1.\label{5.1}
   \end{equation}
The space $T(A_\infty(\oR^4))$ consists of all terminating
sequences of the form $\mathbf f=(f_0,f_1,\dots)$, where $f_n\in
A_\infty(\oR^{4n})$ and only a finite number of  $f_n$'s are
different from zero. It is an involutory algebra under the
multiplication
\begin{equation}
(\mathbf f\otimes\mathbf  g)_n=\sum_{k=0}^nf_k\otimes g_{n-k}
\label{5.2}
   \end{equation}
and with the involution
\begin{equation}
f_n^\dagger(z_1,\dots,z_n)=\overline{f_n(\bar z_n,\dots,\bar
z_1)}. \label{5.3}
   \end{equation}
Operations~\eqref{5.2} and \eqref{5.3} are continuous in the
natural topology of the direct sum. Furthermore,
$T(A_\infty(\oR^4))$ is a nuclear  LF-space.

By  conditions $(a.2)$ and $(a.3)$, the algebra
$T(A_\infty(\oR^4))$ can be equipped   with the positive
semidefinite hermitian form
\begin{equation}
s(\mathbf f,\mathbf g)=\sum_{k,m\geq0}\left({\mathcal
W}_{k+m},f^\dagger_k\otimes g_m\right), \label{5.4}
   \end{equation}
which  defines a nondegenerate inner product $\langle
\cdot,\cdot\rangle$ on the quotient space $T(A_\infty(\oR^4))/\ker
s$. Completing the latter with respect to the corresponding  norm
$\|\cdot\|$, we obtain a Hilbert space $\mathcal H$. There is a
natural continuous linear map $T(A_\infty(\oR^4))\to\mathcal H$.
We denote by $D$ its image, which is a dense subspace of $\mathcal
H$, and by $\Psi_{\mathbf f}$ the image of  $\mathbf f$ in
$\mathcal H$. Since every space $A_\infty(\oR^{4n})$ is separable,
so is $\mathcal H$.

From  $(a.4)$ it follows that $\mathbf f\in \ker
s\Rightarrow\mathbf f_{(a,\Lambda)}\in \ker s$. Therefore, the
action of ${\mathcal P}^\uparrow_+$ in $T(A_\infty(\oR^4))$
induces a linear representation $(a,\Lambda)\to U(a,\Lambda)$ of
this group in $D$, defined by
\begin{equation}
U(a,\Lambda)\Psi_{\mathbf f}=\Psi_{\mathbf f_{(a,\Lambda)}}.
\label{5.5}
   \end{equation}
The condition $(a.4)$ implies also that every operator
$U(a,\Lambda)$ is isometric. Besides, it is bijective and hence
can be extended by continuity to a unitary operator on the whole
of $\mathcal H$.  The map $(a,\Lambda)\to \langle \Psi_{\mathbf
f},U(a,\Lambda)\Psi_{\mathbf f}\rangle=s(\mathbf f,\mathbf
f_{(a,\Lambda)})$ is continuous for each $\mathbf f$, because
${\mathcal P}^\uparrow_+$ acts continuously in every space
$A_\infty(\oR^{4n})$. Therefore, $\|\Psi_{\mathbf
f}-U(a,\Lambda)\Psi_{\mathbf f}\|\to 0$ as $(a,\Lambda)\to (0,I)$.
Using the unitarity of $U(a,\Lambda)$ in the same manner as
in~\cite{SW}, we deduce that this operator is continuous in
$(a,\Lambda)$ on a general $\Psi\in \mathcal H$.

We denote by $\Psi_0$ the image of $(1,0,0,\dots)$ in $\mathcal
H$. Clearly, $U(a,\Lambda)\Psi_0=\Psi_0$ for all $(a,\Lambda)$.
There is no other translation invariant state in $\mathcal H$.
Indeed, assume the converse, that such a state $\Phi_0$ exists.
Without loss of generality we can also assume that $\langle
\Phi_0,\Psi_0\rangle=0$ and $\|\Phi_0\|=1$. Since $D$ is dense in
$\mathcal H$, for any $\epsilon>0$ there is a vector
$\Psi_{\mathbf f}$ such that $\|\Psi_{\mathbf
f}-\Phi_0\|<\epsilon$. Denoting  $\Psi_{\mathbf f}-\Phi_0$ by
$\Phi$, we have
\begin{multline}
\langle \Psi_{\mathbf f}, U(\lambda a, I)\Psi_{\mathbf
f}\rangle=\langle \Phi_0,
U(\lambda a, I)\Phi_0\rangle\\
+\langle \Phi, U(\lambda a, I)\Phi_0\rangle+\langle \Phi_0,
U(\lambda a, I)\Phi\rangle+\langle \Phi, U(\lambda a,
I)\Phi\rangle, \label{5.6}
   \end{multline}
where $\langle \Phi_0, U(\lambda a, I)\Phi_0\rangle\equiv 1$.
Applying the Schwarz inequality and using the unitarity of
$U(\lambda a, I)$, we see that the sum of last three terms on the
right-hand side of~\eqref{5.6} is bounded in absolute value by
$2\epsilon +\epsilon^2$. On the other hand, if $a$ is spacelike
and $\lambda\to\infty$, then from the cluster decomposition
property $(a.6)$ it follows that
\begin{multline}
 \langle \Psi_{\mathbf f}, U(\lambda a,
I)\Psi_{\mathbf f}\rangle\equiv \sum_{k,m\geq0}\left({\mathcal
W}_{k+m},f^\dagger_k\otimes (f_m)_{(\lambda a,I)}\right)\\ \to
\sum_{k,m\geq0}({\mathcal W}_k,f^\dagger_k)({\mathcal W}_m,
f_m)=\langle \Psi_{\mathbf f}, \Psi_0\rangle\langle \Psi_0,
\Psi_{\mathbf f}\rangle, \label{5.7}
  \end{multline}
where $|\langle \Psi_{\mathbf f}, \Psi_0\rangle|=|\langle \Phi,
\Psi_0\rangle|<\epsilon$. Thus, we obtain a contradiction for
sufficiently small~$\epsilon$.

Since $U(a,I)$ is  unitary and continuous, it can be written as
$U(a,I)=e^{ia^\nu P_\nu}$, where  $P_\nu$ are commuting
self-adjoint operators. From  $(a.5)$ it follows that the spectrum
of the energy-momentum operator  $P$ is contained in the closed
forward cone ${\overline \oV}^+$. This can be proved by the
standard arguments~\cite{SW,J} because the Fourier transform of
$A_\infty$ contains the space $\mathcal D$ of smooth functions of
compact support. Namely, the spectral representation of $U(a,I)$
shows that the statement about the spectrum of $P$ amounts to
saying that $\int{\rm d}a \rho(a)U(a,I)=0$ for all $\hat\rho$
belonging to $\mathcal D(\oR^4)$ and supported in the complement
of ${\overline \oV}^+$. Let $\mathbf f$ be such that only one of
its components $f_k$ is nonzero and let $\mathbf g$ be such that
only $g_m\ne0$. Then we have
\begin{equation}
\langle \Psi_{\mathbf f}, \int\!{\rm d}a\,
\rho(a)U(a,I)\Psi_{\mathbf g}\rangle=\left(\mathcal
W_{k+m},f^\dagger_k\otimes \int\!{\rm d}a \,\rho(a)
(g_m)_{(a,I)}\right). \label{5.8}
   \end{equation}
The Fourier transform of $\int\!{\rm d}a \,\rho(a) (g_m)_{(a,I)}$
is equal to ${\hat \rho}(p_1+\dots+p_m){\hat g}_m(p_1,\dots,p_m)$.
In terms of the functionals $\mathcal W_n$,   condition $(a.5)$
means that  $\supp\hat{\mathcal W}_n(p_1,\dots,p_n)$ is contained
in the  set defined by $p_j+\dots+p_n\in -{\overline \oV}^+$,
$j=1,2,\dots,n$. Because $({\mathcal W}_n,
h)=(2\pi)^{4n}(\hat{\mathcal W}_n, \hat h(-\cdot))$, we conclude
that  matrix element~\eqref{5.8}   vanishes if $\supp\hat\rho$
does not intersect ${\overline \oV}^+$, which proves the
statement.

For each  $h\in A_\infty(\oR^4)$, we define a linear operator
$\varphi(h)$ on $D$ by
\begin{equation}
\varphi(h)\Psi_{\mathbf f}=\Psi_{h\otimes\mathbf  f},\quad
\mbox{where} \quad h\otimes\mathbf f=(0,hf_0,h\otimes f_1,h\otimes
f_2,\dots). \label{5.9}
   \end{equation}
It is well defined because if $\mathbf f\in \ker s$, then
$h\otimes\mathbf f\in \ker s$. Indeed, we have $(h\otimes\mathbf
f)^\dagger=\mathbf f^\dagger\otimes h^\dagger$ and hence
$s(h\otimes\mathbf f,\mathbf g)=s(\mathbf f, h^\dagger\otimes
\mathbf g)$ by  definition~\eqref{5.4}. By the same argument,
$$
\langle \Phi, \varphi(h)\Psi\rangle= \langle
\varphi(h^\dagger)\Phi, \Psi\rangle \qquad\mbox{for all}\quad
\Phi,\Psi\in D\quad\mbox{and}\quad h\in A_\infty(\oR^4),
$$
i.e., the field $\varphi$ is hermitian. Clearly,
$\varphi(h)D\subset D$. Since $A_\infty(\oR^4)^{\otimes n}$ is
dense in $A_\infty(\oR^{4n})$, the vector $\Psi_0$  is cyclic for
$\varphi$. All the matrix elements $\langle \Phi,
\varphi(h)\Psi\rangle$, where $\Phi,\Psi\in D$, can be  expressed
in terms of the functionals $\mathcal W_n$ with fixed  $(n-1)$
arguments and hence they belong to $A'_\infty(\oR^4)$. From the
relation $(h\otimes \mathbf
f)_{(a,\Lambda)}=h_{(a,\Lambda)}\otimes\mathbf f_{(a,\Lambda)}$
and  definition~\eqref{5.5}, it follows that
$$
U(a,\Lambda)\varphi(h)U(a,\Lambda)^{-1}=\varphi(h_{(a,\Lambda)}).
$$

Suppose now that $\tilde{\mathcal H}$, $\tilde U(a,\Lambda)$, and
$\tilde\varphi$ define a field theory with a cyclic vacuum state
$\tilde\Psi_0$ and with the same expectation values. Let $\tilde
D_0$ be the vector subspace spanned by $\tilde\Psi_0$ and all
vectors of the form
\begin{equation}
\underbrace{\tilde\varphi(f)\tilde\varphi(g)\dots
\tilde\varphi(h)}_n\tilde\Psi_0, \label{5.10}
   \end{equation}
where $n=1,2,\dots$ and all test functions are in
$A_\infty(\oR^4)$. We assert that the multilinear map taking each
$n$-tuple
\begin{equation}
(f,g,\dots, h)\in\underbrace{A_\infty(\oR^4)\times\dots
A_\infty(\oR^4)}_n
\notag
\end{equation}
 to  vector~\eqref{5.10} is separately
continuous. We consider it as a function of one of variables, say,
of $g$ with all other variables held fixed. Let $\tilde\Phi$ be an
arbitrary element of $\tilde{\mathcal H}$. Since $\tilde D_0$ is
dense in $\tilde{\mathcal H}$, there exists a sequence
$\tilde\Phi_\nu\in \tilde D_0$ such that
$\tilde\Phi_\nu\to\tilde\Phi$. The sequence of  continuous linear
forms $g\to
\langle\tilde\Phi_\nu,\tilde\varphi(f)\tilde\varphi(g)\dots
\tilde\varphi(h)\tilde\Psi_0\rangle$ converges pointwise to the
form $g\to \langle\tilde\Phi,\tilde\varphi(f)\tilde\varphi(g)\dots
\tilde\varphi(h)\tilde\Psi_0\rangle$ and  the latter is also
continuous by the uniform boundedness principle which is
applicable because $A_\infty(\oR^4)$ is a Fr\'echet space. It
follows that the map $A_\infty(\oR^4)\to\tilde{\mathcal H}\colon
g\to
\tilde\varphi(f)\tilde\varphi(g)\dots\tilde\varphi_1(h)\tilde\Psi_0$
is weakly continuous. But for the class of Fr\'echet spaces, the
weak continuity of a linear map is equivalent to the continuity in
their original topology (see~\cite{Sch}, Sec.~IV.7.4), which
proves our assertion. Using  the kernel theorem for $A_\infty$, we
conclude that the multilinear map under consideration can be
uniquely extended to a continuous linear map
$A_\infty(\oR^{4n})\to \tilde{\mathcal H}$, which gives an exact
meaning to the formal expression
\begin{equation}
\int {\rm}dx_1\dots {\rm}dx_n
f(x_1,\dots,x_n)\prod_{i=1}^n\tilde\varphi(x_i)\tilde\Psi_0,\qquad
\mbox{where} \quad f\in A_\infty(\oR^{4n}),\notag
\end{equation}
and also to the vectors
\begin{equation}
\tilde\Psi_{\mathbf f}=f_0\tilde\Psi_0+\int {\rm}dx
f_1(x)\tilde\varphi(x)\tilde\Psi_0 +\int {\rm}dx_1 {\rm}dx_2
f_2(x_1,x_2)\tilde\varphi(x_1)\tilde\varphi(x_2)\tilde\Psi_0+\dots.\notag
\end{equation}
Let $V$ be  the map  taking  $\Psi_{\mathbf f}\in\mathcal H$ to
$\tilde\Psi_{\mathbf f}\in\tilde{\mathcal H}$. This map is well
defined because $\mathbf f\in \ker s$ implies that
$\langle\tilde\Phi,\tilde\Psi_{\mathbf f}\rangle=0$ for all
$\tilde\Phi\in \tilde D_0$ and hence $\tilde\Psi_{\mathbf f}=0$.
From the equality  of the  expectation values in the two theories,
we also deduce  that the operator  $V$ is isometric. Since $D$ is
dense in $\mathcal H$ and $\tilde D_0$  is dense in
$\tilde{\mathcal H}$, this operator can be extended by continuity
to a unitary operator from  $\mathcal H$ onto $\tilde{\mathcal
H}$. We have the chain of equalities
$$
V\varphi(h)\Psi_{\mathbf f}=V\Psi_{h\otimes \mathbf
f}=\tilde\Psi_{h\otimes \mathbf
f}=\tilde\varphi(h)\tilde\Psi_{\mathbf f}=\tilde\varphi
(h)V\Psi_{\mathbf f},
$$
and hence $V\varphi(h)V^{-1}=\tilde\varphi(h)$. The transformation
law of $\tilde\varphi$ under the Poincar\'e group implies that
$\tilde U(a,\Lambda)\tilde\Psi_{\mathbf f}=\tilde\Psi_{\mathbf
f_{(a,\Lambda)}}$. Therefore, an analogous chain of equalities
gives $U(a,\Lambda)=V\tilde U(a,\Lambda)V^{-1}$, which completes
the proof of Theorem~1.

We now turn to the proof of Theorem~2. Let $\varphi$ be the field
constructed above. First we show that under condition $(a.7.1)$,
the vector-valued generalized function $\varphi(g)\Psi_0$, $g \in
A_\infty(\oR^4)$, has a continuous extension to the space
$A_{\ell/2}(\oR^4)$. It follows from $(a.7.1)$ that there are
$l<\ell$ and $N$ such that
$\|W_2\|_{l,N}\equiv\sup_{\|f\|_{l,N}\leq1}|(W_2,f)|<\infty$.
Using~\eqref{3.4}, the triangle inequality for the norm
$|\zeta|=\max_j|\zeta_j|$, and the analyticity of $g$, we obtain
the estimate
\begin{multline}
\|\varphi(g)\Psi_0\|^2 =(\mathcal W_2,g^\dagger\otimes
g)=\left(W_2,\int \overline{g(x+\xi)}g(x){\rm}dx \right)\leq
\|W_2\|_{l,N} \left\|\int
 \overline{g(x+\xi)}g(x){\rm}dx\right\|_{l,N}\\
\leq \|W_2\|_{l,N}
\sup_{|\eta|<l}\sup_\xi\,(1+|\xi+i\eta|)^{N}\left|\int
\overline{g(x+ \xi-i\eta)} g(x){\rm}dx\right|\\
\leq (1+l)^N\|W_2\|_{l,N}  \sup_{|\eta|<l}\sup_\xi\,\int
(1+|x+\xi|)^{N}|g(x+ \xi-i\eta/2)|\,(1+|x|)^{N} |g(x-i\eta/2)|{\rm}dx\\
\leq 2^{2N+5}(1+l)^N\|W_2\|_{l,N}\|g\|_{l/2,N}\|g\|_{l/2,N+5}\int
\frac{{\rm}dx}{(1+|x|)^5}\leq C \|g\|_{l/2,N+5}^2,
 \notag
\end{multline}
which demonstrates the existence of the desired extension. Next we
consider the vector-function
 \begin{equation}
\Psi(g^{(r)})=\int {\rm}dx_1\dots {\rm}dx_n
g(x_1,x_1-x_2\dots,x_{n-1}-x_n)\prod_{i=1}^n\varphi(x_i)\Psi_0,
 \label{5.11}
\end{equation}
where $n\geq 2$. The reasoning after formula~\eqref{5.10} shows
that it is well defined on the space $A_\infty(\oR^{4n})$, for
which the map $g\to g^{(r)}$ is an automorphism. We express the
squared norm of  vector~\eqref{5.11} in terms of $W_{2n}$'s and
denote by $\|g\|_{l/2,l,N}$ the norm of $g$ in the space
$A_{\ell/2,\ell,N}(\oR^4\times\oR^{4(n-1)})$. Proceeding along the
same lines as above and using in addition the elementary
inequality $1+|\xi|\leq (1+|\xi_n|)(1+\max_{j\ne n}|\xi_j|)$, we
get
\begin{multline}
\|\Psi(g)\|^2=\left(W_{2n}, \int  \overline{g(x+\xi_n,
-\xi_{n-1},\dots
-\xi_1)} g(x,\xi_{n+1},\dots,\xi_{2n-1}){\rm d}x \right)\\
\leq
2^{4N+5}(1+l)^N\|W_{2n}\|_{l,N}\|g\|_{l/2,l,2N}\|g\|_{l/2,l,2N+5}\int
\frac{{\rm d}x}{(1+|x|)^{5}}\leq C'\|g\|_{l/2,l,2N+5}^2
 \notag
\end{multline}
and conclude that the linear map  $g\to\Psi(g^{(r)})$ has a
continuous extension to the space
$A_{\ell/2,\ell}(\oR^4\times\oR^{4(n-1)})$. It is important that
the extension is unique because $A_\infty(\oR^{4n})$ is dense in
this space, as already noted in Remark~1. If we replace $\Psi_0$
in~\eqref{5.11} with an arbitrary $\Psi_{\mathbf f}\in D$, then
the resulting vector-function also has a unique  continuous
extension to $A_{\ell/2,\ell}(\oR^4\times\oR^{4(n-1)})$. Indeed,
it is a finite sum of vector functions of the previous form but
with the difference that $g$ is now replaced by $g\otimes f_m$,
where $f_m\in A_\infty(\oR^{4m})$. If $g\in
A_{\ell/2,\ell}(\oR^4\times \oR^{4(n-1)})$, then $g\otimes f_m\in
A_{\ell/2,\ell}(\oR^4\times\oR^{4(n+m-1)})$ and the map $g\to
g\otimes f_m$   is continuous. Thus, from $(a.7.1)$ it follows
that every  monomial $\prod_{i=1}^n\varphi(x_i)$, $n\geq2$, has a
unique continuous extension to an operator-valued generalized
function defined on $A^{(r)}_\ell(\oR^{4n})$. Conversely, in any
field theory with such a property, the $n$-point vacuum
expectation value $W_n$ is well defined on $A_\ell(\oR^{4(n-1)})$
by the formula
$$
(W_n,f)=\left\langle\Psi_0,\int {\rm}dx_1\dots {\rm}dx_n
g_1(x_1)f(x_1-x_2\dots,x_{n-1}-x_n)\prod_{i=1}^n\varphi(x_i)\Psi_0
\right\rangle,
$$
where $g_1$ is any element of $A_{\ell/2}(\oR^4)$ such that $\int
g_1(x){\rm}\,dx =1$. Theorem~2 is proved.

{\it Proof of Theorem~3}: Combining Theorem~7 and Corollary of
Theorem~2 we infer that the matrix element
$\langle\Phi,[\varphi(x), \varphi(x')]_-\Psi\rangle$ is well
defined as a continuous linear functional on
$A_{\ell/2}(\oR^{4\cdot2})$ for any fixed $\Phi,\Psi\in D_\ell$.
For brevity, we denote this functional by $u_{\Phi,\Psi}$. Let
\begin{equation}
\Phi=\int {\rm}dx_1\dots {\rm}dx_{k-1}
\overline{h(x_{k-1},x_{k-1}-x_{k-2},\dots,x_2-x_1)}\,
\varphi(x_{k-1})\cdots \varphi(x_1)\Psi_0
\label{5.12}
\end{equation}
and
\begin{equation} \Psi=\int
{\rm}dx_{k+2}\dots {\rm}dx_n
g(x_{k+2},x_{k+2}-x_{k+3},\dots,x_{n-1}-x_n)\,\varphi(x_{k+2})\cdots
\varphi(x_n)\Psi_0,
\label{5.13}
\end{equation}
where  $k\geq 2$, $n\geq k+2$, $h\in
A_{3\ell/2,\ell}(\oR^4\times\oR^{4(k-2)})$, and $g\in
A_{3\ell/2,\ell}(\oR^4\times\oR^{4(n-k-2)})$. Let $f(x,x')$ belong
to $A_{\ell/2}(\oR^{4\cdot2})$. We introduce the notation
$\xi_{k-1}=x_{k-1}-x$, $\xi_k= x-x'$, $\xi_{k+1}=x'-x_{k+2}$. The
number $(u_{\Phi,\Psi},f)$ is equal to the value of
functional~\eqref{2.3} at the function
\begin{multline}
F(\xi_1,\dots,\xi_{n-1})=\\\int {\rm d} x \,h(x+\xi_{k-1},\,
-\xi_{k-2},\dots, -\xi_1)\,\, f(x, x-\xi_k)
g(x-\xi_k-\xi_{k+1},\,\xi_{k+2},\dots,\xi_{n-1}).
 \label{5.14}
\end{multline}
From $(a.7.2)$ it follows that  functional~\eqref{2.3} is bounded
in one of the norms of the space $A_\ell(V_{(k)})$ with some
indices $l<\ell$, $N$. Proceeding in a manner similar to that used
in proving Theorem~2 and taking into account that the plane of
integration in~\eqref{5.14} may be shifted  within the analyticity
domain, we obtain
\begin{multline}
|(u_{\Phi,\Psi},f)|\leq C\,\sup_{|\eta|\leq l}\,\sup_{\xi\in
V_{(k)}}(1+|\xi|)^N|F(\xi+i\eta)|
\\
\leq C\sup_{|\eta|\leq l}\,\sup_{\xi\in V_{(k)}}(1+\max_{j<
k-1}|\xi_j|)^N(1+|\xi_{k-1}|)^N (1+|\xi_k|)^N (1+|\xi_{k+1}|)^N
(1+\max_{j> k+1}|\xi_j|)^N|
F(\xi+i\eta)|\\
 \leq C' \|h\|_{3l/2,l,2N}\|g\|_{3l/2,l,2N}
 \sup_{|\eta_k|\leq l}\sup_x\sup_{\xi_k^2\geq 0}(1+|x|)^{2N+5}
(1+|x-\xi_k|)^{2N}|f(x+i\eta_k/2,x-\xi_k-i\eta_k/2)|\\
\leq
C^{\prime\prime}\|h\|_{3l/2,l,2N}\|g\|_{3l/2,l,2N}\|f\|_{\oW,l/2,4N+5}.
\notag
\end{multline}
Clearly, a similar estimate holds if $\Psi_0$ in~\eqref{5.12},
\eqref{5.13} is changed for an arbitrary vector in $D$  and also
in the event that $\Phi=\Psi_0$, or $\Psi=\Psi_0$, or
$\Phi=\Psi=\Psi_0$. Therefore, every functional $u_{\Phi,\Psi}$,
where $\Phi,\Psi\in D_\ell$, has a continuous extension to the
space $A_{\ell/2}(\oW)$. This completes the proof of the
reconstruction theorem.

\section{Two formulations of quasilocality}

In this section, we show that the property  stated in Theorem~3
faithfully enough reproduces the initial property $(a.7.2)$ of the
Wightman functionals. Namely, if the   commutator
$[\varphi(x),\varphi(x')]_-$ in a  field theory has this property,
then  functional~\eqref{2.3} composed of the $n$-point vacuum
expectation values of  $\varphi$ extends continuously to the space
$A_{2\ell}(V_{(k)})$. We shall use a slightly different version of
Lemma~1.

{\bf  Lemma~2:} {\it Let $L$ be a sequentially dense subspace of a
locally convex space  $F$ and let $G_1,\dots, G_n$ be barrelled
spaces. Then every  separately continuous multilinear form $\mu$
defined on $L\times G_1\times\dots G_n$ has a unique extension to
a separately continuous multilinear form on $F\times
G_1\times\dots G_n$.

Proof:} For each fixed  $g_j\in G_j$, $j=1,\dots,n$, the linear
form $f\to\mu(f,g_1,\dots,g_n)$ can be uniquely extended to  $F$
by continuity. Letting $\hat{\mu}$ denote this extension, we have
to show that it is linear and continuous in every $g_j$ for any
fixed $f\in F$ and $g_i\in G_i$, $i\ne j$. We  set  $j=1$  without
loss of generality. Choose a sequence $f_\nu\in L$ such that
$f_\nu\to f$ in $F$ and denote by $f^{(g_2,\dots,g_n)} _\nu$ the
corresponding elements of $G'_1$ defined by
$f^{(g_2,\dots,g_n)}_\nu(g_1)=\mu(f_\nu, g_1,g_2,\dots,g_n)$. The
sequence $f^{(g_2,\dots,g_n)} _\nu$ converges pointwise  on $G_1$
to the functional $f^{(g_2,\dots,g_n)}(g_2)=\hat{\mu}(f,
g_1,g_2,\dots,g_n)$ and hence this functional is linear and
continuous because $G_1$ is barrelled. The lemma is thus proved.

Let $\varphi$ be a scalar field with test functions in
$A_\infty(\oR^4)$. Suppose that every  monomial
$\prod_{i=1}^n\varphi(x_i)$ has a continuous extension to the
space of all functions of  form~\eqref{2.5}. Then, as shown in
Sec.~II, the operators $\varphi(f)$, $f\in A_{\ell/2}(\oR^4)$, are
well defined and act continuously on the linear span of the vacuum
state $\Psi_0$ and  all vectors of the form
\begin{equation}
\int {\rm}dx_1\dots {\rm}dx_n\,
g(x_1,x_1-x_2\dots,x_{n-1}-x_n)\prod_{i=1}^n\varphi(x_i)\Psi_0,\quad
n\geq1, \label{6.1}
\end{equation}
where $g$ ranges over the space
$A_{3\ell/2,\ell}(\oR^4\times\oR^{4(n-1)})$. Suppose that the
matrix element $\langle\Phi,[\varphi(x),
\varphi(x')]_-\Psi\rangle$ has a continuous extension to
$A_{\ell/2}(\oW)$ for any states $\Phi$ and $\Psi$ of
form~\eqref{6.1}. Then {\it a fortiori} it has a unique continuous
extension  to the space $\mathcal A_{\ell/2}(\oW)$ introduced in
Sec.~IV. We take $\Phi$ and $\Psi$ to be the vectors that are
defined by~\eqref{5.12} and~\eqref{5.13} with the following choice
of test functions:
\begin{equation}
 h(x_{k-1},-\xi_{k-2},\dots,-\xi_1)=h_1(x_{k-1})\mathbf
h(\xi_1,\dots,\xi_{k-2}),\quad h_1\in A_{3\ell/2}(\oR^4),\,\,
\mathbf h\in A_\ell(\oR^{4(k-2)}), \notag
\end{equation}
and
\begin{equation}
g(x_{k+2},\xi_{k+2},\dots,\xi_{n-1})=g_1(x_{k+2})\mathbf
g(\xi_{k+2},\dots,\xi_{n-1}),\quad g_1\in A_{3\ell/2}(\oR^4),\,\,
\mathbf g\in A_\ell(\oR^{4(n-k-2)})\notag.
 \notag
\end{equation}
By Theorem~8 and  Lemma~2, which is applicable because any
Fr\'echet space is barrelled, the matrix element under
consideration extends uniquely to a trilinear separately
continuous form $\mu(\mathbf h,f,\mathbf g)$ on the space
$A_\ell(\oR^{4(k-2)})\times F\times A_\ell(\oR^{4(n-k-2)})$, where
$$
F=A_{3\ell/2}(\oR^4)\hotimes \mathcal A_{\ell/2}(\oW) \hotimes
A_{3\ell/2}(\oR^4)=\mathcal
A_{3\ell/2,\ell/2,3\ell/2}(\oR^4\times\oW\times\oR^4).
$$
Let $f_0$ be a fixed element of $A_\infty(\oR^4)$ and let $\mathbf
f\in \mathcal A_{2\ell,\ell,2\ell}(\oR^4\times\oV\times\oR^4)$.
Then the function
\begin{equation}
f(x_{k-1},x,x',x_{k+2})=f_0\left(\frac{x+x'}{2}\right)\mathbf
f(x_{k-1}-x, x-x',x'-x_{k+2})
 \label{6.2}
\end{equation}
belongs to  $F$ and the map $\iota\colon \mathcal
A_{2\ell,\ell,2\ell}(\oR^4\times\oV\times\oR^4)\to F\colon \mathbf
f\to f$ is continuous. Using Theorem~8 again, we infer that the
trilinear form $\mu(\mathbf h, \iota(\mathbf f),\mathbf g)$
extends uniquely  to a continuous linear functional on the space
\begin{multline}
A_\ell(\oR^{4(k-2)})\hotimes \mathcal
A_{2\ell,\ell,2\ell}(\oR^4\times\oV\times\oR^4)\hotimes
A_\ell(\oR^{4(n-k-2)})\\=\mathcal
A_{\ell,2\ell,\ell,2\ell,\ell}(\oR^{4(k-2)}\times
\oR^4\times\oV\times\oR^4\times \oR^{4(n-k-2)}). \label{6.3}
\end{multline}
Now we consider  functional~\eqref{2.3} with $W_n$ taken to be the
$n$-point Wightman function of $\varphi$. If $\mathbf h \in
A_\infty(\oR^{4(k-2)})$, $\mathbf f\in A_\infty(\oR^{4\cdot 3})$,
$\mathbf g \in A_\infty(\oR^{4(k-n-2)})$, and $\int f_0(X){\rm
d}X=1$,  then $\mu(\mathbf h, \iota(\mathbf f),\mathbf g)$
coincides with the value of functional~\eqref{2.3} at the test
function $\mathbf h\otimes \mathbf f\otimes\mathbf g$. The linear
span of all functions of this form is dense in space~\eqref{6.3},
and we conclude that functional~\eqref{2.3} has a unique
continuous extension to this space. {\it A fortiori} it can be
continuously extended to $\mathcal A_{2\ell}(V_{(k)})$ and, by the
Hahn-Banach theorem, to $A_{2\ell}(V_{(k)})$.

\section{Concluding remarks}

An interesting feature of the reconstruction theorem established
in this paper is the necessity of using the extended domain
$D_\ell\subset \mathcal H$ for the operator realization of the
quasilocality condition. It cannot be replaced by the invariant
domain $D$ spanned by all vectors of the form
\begin{equation}
 \int {\rm}dx_1\dots {\rm}dx_n
f(x_1,\dots,x_n)\prod_{i=1}^n\varphi(x_i)\Psi_0,\label{7.1}
\end{equation}
where $f\in A_\infty(\oR^{4n})$, because such a simplified
formulation is not equivalent to the initial assumption $(a.7.2)$
for the Wightman functionals. At this point, there is a
significant difference to the situation in local
QFT~\cite{SW,J,BLOT}, where the invariant domain $D_0$ spanned by
all vectors $\varphi(f)\varphi(g)\dots \varphi(h)\Psi_0$, with
$f,g,\dots,h$ ranging over $\mathcal D(\oR^4)$, is large enough
for formulating the microcausality condition. Then this condition
is also satisfied for the field commutator acting on any vector of
form~\eqref{7.1} with $f\in S(\oR^{4n})$, because $\mathcal
D(\oR^4)^{\otimes n}$ is dense in $S(\oR^{4n})$ and the
distributions supported in a given closed set form a closed set in
the space  of tempered distributions. On the contrary, the
subspace of $A'_\ell(\oR^d)$ consisting of those  functionals that
are carried by a closed set $M\subset\oR^d$ is everywhere dense in
$A'_\ell(\oR^d)$. Indeed, if this were not the case, then by the
Hahn-Banach theorem there would exist a nontrivial function $f\in
A_\ell(\oR^d)$ such that $(u,f)=0$  for all $u$ in this subspace
because $A_\ell(\oR^d)$ is reflexive. In particular, $(\delta_z,
f)=f(z)=0$ for all $z\in\tilde M^\ell$, but this contradicts the
analyticity of $f$.

Some examples of nonlocal but quasilocalizable fields were
discussed in~\cite{I,BN}. The simplest model of this kind is the
normal ordered Gaussian function $:e^{g\phi^2}:(x)$ of a free
neutral scalar field $\phi$. As shown in~\cite{I}, the vacuum
expectation values
\begin{equation}
{\mathcal W}_n(x_1,\dots x_n)=\langle\Psi_0,
:e^{g\phi^2}:(x_1)\,\ldots :e^{g\phi^2}:(x_n) \Psi_0\rangle,
\label{7.2}
\end{equation}
calculated by the Wick theorem, satisfy  conditions $(a.7.1)$ and
$(a.7.2)$ with  $\ell=\sqrt{2g/3}$. Therefore, the field
$:e^{g\phi^2}:(x)$ has the properties established by Theorems~2
and 3. Moreover, these properties are characteristic of any field
$\varphi$ expressed as
\begin{equation}
\varphi(x)=\sum_{r=0}^\infty\frac{d_r}{r!}:\phi^r:(x), \label{7.3}
\end{equation}
where the coefficients $d_r$  satisfy the inequality $d_r^2\leq C
(2g)^r r!$. It is worth  noting that  these fields can be
implemented directly in the Fock space $\mathcal H_0$ of the
initial field $\phi$ without appealing to the reconstruction
theorem. A simple method of analyzing the conditions for
convergence  of  infinite series in Wick powers of a free field
has been proposed in~\cite{SS1}. This method systematically uses
the analyticity properties of Wightman functions and is well
applicable to the nonlocalizable power series.

Conditions $(a.7.1)$ and $(a.7.2)$ can  be  weakened by assuming
that the nonlocality parameter increases with $n$. Such a
modification would be appropriate if a future investigation of
physically relevant models (related, e.g., to string theory) would
give sufficient grounds. For instance, it may be suggested that
for any integer $k\geq 1$, there is a positive number $\gamma(k)$
such that the following requirements are fulfilled.
\begin{enumerate}
\item[$(a.7.1)'$] Every functional  $\mathcal W_n$, $n>1$, has a
continuous extension to each of the spaces
$$
A_{\ell\gamma(k-1)}(\oR^{4(k-1)})\hotimes
A_{\ell/2}(\oR^{4\cdot2}) \hotimes
A_{\ell\gamma(n-k-1)}(\oR^{4(n-k-1)}),\quad 1\leq k\leq n-1.
$$
\item[$(a.7.2)'$] For any $n\geq2$ and $k\leq n-1$, the difference
$$
\mathcal W_n(x_1,\dots,x_k,x_{k+1},\dots, x_n) - \mathcal
W_n(x_1,\dots,x_{k+1},x_k,\dots, x_n)
 $$
has a continuous extension to the space
$$A_{\ell\gamma(k-1)}(\oR^{4(k-1)})\hotimes A_{\ell/2}(\oW)
\hotimes A_{\ell\gamma(n-k-1)}(\oR^{4(n-k-1)}).
$$
\end{enumerate}
It follows from $(a.7.1)'$  that the operator-valued generalized
function $f\to \varphi(f)$ constructed by Theorem~1 extends
continuously to the space $A_{\ell/2}(\oR^4)$. Moreover, this
condition is equivalent to the fact that every  monomial
$\prod_{j=1}^{n+1}\varphi(f_j)$ extends continuously to
$A_{\ell/2}(\oR^4)\hat\otimes A_{\ell \gamma(n)}(\oR^{4n})$. The
operator-valued generalized function $\varphi(f)$, $f\in
A_{\ell/2}(\oR^4)$, is thereby defined on the linear span
$D'_\ell$ of all vectors of the form
\begin{equation}
\int {\rm}dx_1\dots {\rm}dx_n
g(x_1,\dots,x_n)\prod_{i=1}^n\varphi(x_i)\Psi,\quad \text{where
$g\in  A_{\ell \gamma(n)}(\oR^{4n})$ and  $\Psi\in D$}.
 \notag
\end{equation}
Condition  $(a.7.2)'$ amounts to    saying that, for any
$\Psi,\Phi\in D'_\ell$, the functional~\eqref{2.6} extends
continuously to the space $A_{\ell/2}(\oW)$.

It is easy to see that conditions $(a.7.1)$ and $(a.7.2)$ imply
$(a.7.1)'$ and $(a.7.2)'$ with $\gamma(k)=k+1/2$. One or the other
of these formulations is preferable according to which variables,
$\xi_j$ or $x_j$, are more convenient to work with.

\medskip

\section*{Acknowledgments}
 This paper was supported in part
by the the Russian Foundation for Basic Research (Grant
No.~09-01-00835) and the Program for Supporting Leading Scientific
Schools (Grant No.~LSS-1615.2008).

\section*{Appendix A: Proof of Theorem~6}

First, we recall Grothendieck's characterization~\cite{Grot2} of
the projective tensor product of complete nuclear spaces. As
above, we use the standard notation $\sigma(F', F)$ for  the weak
topology on the dual of $F$. When $F'$ is provided with this
topology, one writes $F'_\sigma$.  Let $\mathcal B(F'_\sigma,
G'_\sigma)$ be the space of separately continuous bilinear forms
on $F'_\sigma\times G'_\sigma$. As shown by Grothendieck  (see
also~\cite{Sch}, Sec.~IV.9.4), if  $F$ and $G$ are complete
locally convex spaces and at least one of them is nuclear, then
$F\mathbin{\hat{\otimes}} G$ can be identified with the space
$\mathcal B_e(F'_\sigma, G'_\sigma)$ equipped with the
 topology $\tau_e$ of biequicontinuous convergence, which is determined
by the set of seminorms
\begin{equation}
p_{U,V}(b)=\sup_{u\in U^\circ, v\in V^\circ}|b(u_,v)|, \notag
\end{equation}
where  $U$ and $V$ range, respectively, over bases of
neighborhoods of 0 in $F$ and $G$. The natural map $\chi\colon
F\times G\to \mathcal B(F'_\sigma, G'_\sigma)$ takes each pair
$(f,g)$ to the bilinear form
$$
 (f\otimes g)(u,v)= (u,f)(v,g),\qquad
   u\in F', v\in G'.
 \eqno{({\rm A}1)}
$$
We now apply this construction to our situation. Let
$\omega_*\colon F\otimes G\to H$  be the linear map associated
with  $\omega$, and let $\hat \omega_*$ be its extension by
continuity to $F\mathbin{\hat{\otimes}} G=\mathcal B_e(F'_\sigma,
G'_\sigma)$. By definition, $\hat\omega_*$ is a unique continuous
map for which the diagram
  \begin{equation}
\xymatrix{
 &F\mathbin{\hat{\otimes}} G
 \ar[dd]^{\hat\omega_*}\\
 F\times G
 \ar[ur]^{\chi}\ar[dr]_{\omega}\\
 &H
}
 \notag
  \end{equation}
is commutative. All we need to do is to show that $\hat \omega_*$
is an algebraic and topological isomorphism. We first prove that
 $\hat \omega_*$ is injective. By definition,
$$
 \hat \omega_*(f\otimes
g)(x,y)=\omega(f,g)(x,y)=f(x)g(y),\qquad f\in F,\, g\in G.
 \eqno{({\rm A}2)}
$$
Let $b\in \ker\hat \omega_*$ and let
$\{b_\gamma\}_{\gamma\in\Gamma}$ be a net in $F\otimes G$ such
that $b_\gamma\to b$ in the topology of $F\mathbin{\hat{\otimes}}
G$. Then $\hat \omega_* (b_\gamma)\to 0$ in  $H$. We define
$\delta_x$ and $\delta_y$ to be the linear functionals on $F$ and
$G$ such that $(\delta_x,f)= f(x)$ and $(\delta_y,g)=g(y)$. Since
the topologies of $F$ and $G$ are not weaker than that of
pointwise convergence, these functionals are continuous and
belong, respectively,  to $F'$ and $G'$.  For the bilinear forms
$b_\gamma$, we have the relation
$$
b_\gamma(\delta_x, \delta_y)=  \hat \omega_*(b_\gamma)(x,y).
 \eqno{({\rm A}3)}
 $$
Indeed, if $b_\gamma=f\otimes g$, then~(A3) follows immediately
from~(A1), (A2), and each element of $F\otimes G$ is a finite sum
of elements of this form. Passing to the limit, we obtain
$b(\delta_x, \delta_y)=0$, because the topology of $\mathcal
B_e(F'_\sigma, G'_\sigma)$ and $H$ is not weaker than the topology
of pointwise convergence. By the Hahn-Banach theorem the sets
$\{\delta_x\in F'\colon x\in X\}$ and $\{\delta_y\in G'\colon y\in
Y\}$ are total\footnote{A subset of a locally convex space $E$ is
called total in $E$ if its linear span is dense in~$E$.} in
$F'_\sigma$ and $G'_\sigma$ because the duals of the latter
coincide with  $F$ and $G$. Therefore, $b$ is identically zero and
we conclude that $\ker \hat \omega_*=0$.

To prove that   $\hat \omega_*$ is surjective, we use that every
complete nuclear space is semi-reflexive  and hence the strong
topology of $G'$ is compatible with the duality between $G$ and
$G'$, see~\cite{Sch}, Sec.~IV.5. Because of this, condition
$(iii)$ implies that, for each fixed $h\in H$, the map $v\to h_v$
is weakly continuous, i.e., is continuous under the topologies
$\sigma(G',G)$ and $\sigma(F,F')$ (ibid, Sec.~IV.7.4). Therefore,
the bilinear form $b_h\colon(u,v)\to (u, h_v)$ belongs to
$\mathcal B(F'_\sigma, G'_\sigma)$. Clearly, we have the identity
$b_h(\delta_x,\delta_y)=h(x,y)$. Considering a net in $F\otimes G$
which converges to $b_h$ in $F\mathbin{\hat{\otimes}}G=\mathcal
B_e(F'_\sigma, G'_\sigma)$ and using again that the topology of
$\mathcal B_e(F'_\sigma, G'_\sigma)$ and $H$ is not weaker than
that of pointwise convergence, we obtain the equality
$\hat\omega_*(b_h)(x,y)=h(x,y)$. Since it holds for all $x$ and
$y$, we conclude that  $\hat\omega_*(b_h)=h$ and  $\omega_*$ is
hence surjective. It remains to prove that the inverse map
$\omega_*^{-1}$ is continuous. We must show that for each
neighborhood $U$ of $0$ in $F$ and  for each neighborhood $V$ of
$0$ in $G$, there is a neighborhood $W$ of $0$ in $H$ such that
$p_{U,V}(b_h)\leq 1$ for all $h\in W$, or equivalently,
$$
\sup_{h\in W,u\in U^\circ, v\in V^\circ}|(u, h_v)|\leq 1.
 \eqno{({\rm A}4)}
 $$
The set $\{h_v\in F\colon  v\in V^\circ\}$ is bounded because the
polar $V^\circ$ is bounded in $G'$ and the map $v\to h_v$ is
continuous for every fixed $h\in H$. Therefore, the family of
continuous linear maps $H\to F\colon h\to h_v$, where $v$ ranges
$V^\circ$, is pointwise bounded. Since $H$ is assumed to be
barrelled, it follows that  this family is equicontinuous (ibid,
Sec.~III.4.2). Thus, there exists a neighborhood $W$ of $0$ in $H$
such that $h_v\in U$ for all $h\in W$ and for all $v\in V^\circ$.
Then (A4) holds by the definition of the polar. This completes the
proof of Theorem~6.

\section*{Appendix B: Two forms of the kernel theorem}

{\bf Theorem~9.}  {\it Let $F$, $G$, and $H$ be Fr\'echet spaces,
and let $\omega$ be a continuous bilinear map from $F\times G$ to
$H$. Suppose that  for each continuous bilinear form $b\colon
F\times G\rightarrow \mathbb{C}$, there is a unique linear
functional $u_b\in H'$ such that $b=u_b\circ\omega$. Then
$$ F\mathbin{\hat{\otimes}}G= H.
\eqno{({\rm B}1)}
 $$
 More  precisely, the continuous extension of the map
 $\omega_*\colon F\otimes G\rightarrow H$ to the completion of the
 projective tensor product  $F\otimes_\pi G$ is an algebraic and
topological isomorphism.

Proof.} The  dual of $F\otimes_\pi G$ is identified with the space
$\mathcal{B}(F,G)$ of all continuous bilinear forms on $F\times G$
(see~\cite{Sch}, Sec.~III.6.2). We denote by $\hat{\omega}_*$ the
continuous extension of $\omega_*$ to $F\mathbin{\hat{\otimes}}G$
and consider the dual map $\hat{\omega}_*^\prime\colon
H'\rightarrow (F\mathbin{\hat{\otimes}} G)'= \mathcal{B}(F,G)$. By
definition,
$$
\hat{\omega}_*^\prime(u)(x\otimes y)=u(\omega(x\otimes
y))=(u\circ\omega)(x,y),\quad u\in H'.
  $$
It follows from our assumptions that the map
$\hat{\omega}_*^\prime$ is bijective. Therefore, $\hat{\omega}_*$
is injective and has a dense image. Moreover,
$F\mathbin{\hat{\otimes}} G$ is a Fr\'echet space (ibid,
Sec.~III.6.3) and hence the image of $\hat{\omega}_*$ is closed
because that of $\hat{\omega}_*^\prime$ is weakly closed (ibid,
Sec.~IV.7.7). Thus,  $\hat{\omega}_*$ is an algebraic isomorphism.
Using the open mapping theorem, we conclude that this map is also
a topological isomorphism, which completes the proof.

{\it Corollary: Let the assumptions of  Theorem~9 be satisfied,
and let $E$ be a locally convex space. Then for every separately
continuous bilinear map $\beta\colon F\times G\rightarrow E$,
there is a unique continuous linear map $u_\beta\colon
H\rightarrow E$ such that $\beta =u_ \beta\circ\omega$.}

Indeed, this statement expresses the category theoretical meaning
of formula~(B1).

\end{document}